\documentclass[10pt,conference]{IEEEtran}
\usepackage{cite}
\usepackage{amsmath,amssymb,amsfonts}
\usepackage{algorithmic}
\usepackage{graphicx}
\usepackage{textcomp}
\usepackage[dvipsnames]{xcolor}

\usepackage{hyperref}
\usepackage{booktabs}
\usepackage{nicematrix}
\usepackage{bbding}
\usepackage{tcolorbox}

\newcommand{\greencheck}{\textcolor{Green}{\Checkmark}}
\newcommand{\redx}{\textcolor{Red}{\XSolidBrush}}

\def\BibTeX{{\rm B\kern-.05em{\sc i\kern-.025em b}\kern-.08em
    T\kern-.1667em\lower.7ex\hbox{E}\kern-.125emX}}
\begin{document}

\title{CrossCodeBench: Benchmarking Cross-Task Generalization of Source Code Models}


\author{
\IEEEauthorblockN{Changan Niu\IEEEauthorrefmark{1},
Chuanyi Li\IEEEauthorrefmark{1},
Vincent Ng\IEEEauthorrefmark{2},
Bin Luo\IEEEauthorrefmark{1}}
\IEEEauthorblockA{\IEEEauthorrefmark{1}State Key Laboratory for Novel Software Technology, Nanjing University, Nanjing\\
Email: niu.ca@outlook.com, \{lcy,luobin\}@nju.edu.cn}
\IEEEauthorblockA{\IEEEauthorrefmark{2}Human Language Technology Research Institute, University of Texas at Dallas, Richardson, Texas, USA\\
Email: vince@hlt.utdallas.edu}
}

\maketitle

\begin{abstract}

Despite the recent advances showing that a model pre-trained on large-scale source code data is able to gain appreciable generalization capability, it still requires a sizeable amount of data on the target task for fine-tuning. And the effectiveness of the model generalization is largely affected by the size and quality of the fine-tuning data, which is detrimental for target tasks with limited or unavailable resources. Therefore, cross-task generalization, with the goal of improving the generalization of the model to unseen tasks that have not been seen before, is of strong research and application value. 

In this paper, we propose a large-scale benchmark that includes 216 existing code-related tasks. Then, we annotate each task with the corresponding meta information such as task description and instruction, which contains detailed information about the task and a solution guide. This also helps us to easily create a wide variety of ``training/evaluation'' task splits to evaluate the various cross-task generalization capabilities of the model. Then we perform some preliminary experiments to demonstrate that the cross-task generalization of models can be largely improved by in-context learning methods such as few-shot learning and learning from task instructions, which shows the promising prospects of conducting cross-task learning research on our benchmark. We hope that the collection of the datasets and our benchmark will facilitate future work that is not limited to cross-task generalization.

\end{abstract}


\section{Introduction}
\label{section:introduction}

The ``pre-train then fine-tune'' paradigm has become the new favorite in software intelligence~\cite{niu2022deep}. Since pre-training tasks can be done with unlabeled data, a model can be pre-trained on large amounts of, easily accessible data, thus obtaining much common sense and linguistic knowledge~\cite{de2021javabert,chirkova2021empirical,karmakar2021pre}. With this knowledge, pre-trained models can achieve better generalizability, which means that it is able to perform better than its ``no pre-training'' counterpart on a wide variety of software engineering (SE) tasks after being fine-tuned on the data of the target task~\cite{feng2020codebert,guo2021graphcodebert,mastropaolo2021t5-learning,ahmad2021plbart,wang2021codet5,niu2022sptcode,guo2022unixcoder}. Therefore, rather than learning models from scratch, the adoption of pre-trained models as backbone for downstream tasks has become a common practice in the field of software intelligence~\cite{guo2021graphcodebert,gotmare2021cascaded,wang2022bridging}.

However, the fine-tuning stage requires updating the weights of a pre-trained model by training on thousands of supervised labels specific to the target task. Therefore, the ``pre-train then fine-tune'' paradigm still relies on the data from the target task. And the effectiveness of transfer learning of the pre-trained model to the target task depends heavily on the size and quality of the fine-tuning data~\cite{niu2022sptcode}. Unfortunately, in practice, we often encounter situations where we need to apply a pre-trained model to a task with very low available data resources. In such cases, we are unable to fine-tune the pre-trained model on sufficient target task data to obtain a fine-tuned model that can be applied to the target task.

Facing the same issue, pioneers in the field of Natural Language Processing (NLP) have proposed many ways to address this issue. Few-shot learning~\cite{radford2019gpt2,yu2018diverse,gao2021making} skips the fine-tuning phase and applies the pre-trained model directly to the target task, without any weight updates. To bridge the gap between the pre-trained model and the target task, few-shot learning gives the model a task description and a few demonstrations (i.e., supervised examples) of the task at inference time. If there is only one positive example, it is also called one-shot learning~\cite{vinyals2016matching}. Moreover, instead of any examples, zero-shot learning gives the model a natural language description of the task~\cite{brown2020gpt3,raffel2020t5}. In the few/one/zero-shot settings, a large-scale pre-trained language model is able to show strong performance on many NLP tasks and benchmarks, in some cases nearly matching the performance of state-of-the-art fine-tuned systems~\cite{radford2019gpt2,brown2020gpt3,raffel2020t5,gao2021making,ye2021crossfit}. In addition, learning from task instructions~\cite{efrat2020turking,weller2020learning,mishra2022cross} adopts task definition, positive and negative examples, where task definition can be seen as a specification of solving the target task.

The SE community also does some explorations. Mastropaolo et al.~\cite{mastropaolo2021t5-learning} and Wang et al.~\cite{wang2021codet5} utilizes multi-task learning~\cite{raffel2020t5,aghajanyan2021muppet} to achieve a better performance on the target task. By learning multiple related tasks simultaneously, multi-task learning aims to make models exploit both task-generic and task-specific information, thereby improving the model's performance on tasks with low available resources. However, multi-task learning favors tasks with significantly larger amounts of data than others, thus requiring sufficient supervised examples of the target task compared to other tasks to guarantee the availability of the model~\cite{gu2018meta,dou2019investigating}. Rather, Prenner and Robbes~\cite{prenner2021making} experiments with several other techniques that promised a possible benefit for small datasets, i.e., active learning, data augmentation, soft labels, self-training and intermediate-task fine-tuning~\cite{phang2018sentence}. They find that soft labels to be more useful, while other methods are relatively more narrowly applicable, less effective, more costly, or inconclusive. Instead of using any data from some target tasks, Guo et al.~\cite{guo2022unixcoder} directly apply the pre-trained model to the code-to-code retrieval task in order to evaluate the performance of code fragment embeddings. Given that there are many very large-scale pre-trained models of source code being proposed (e.g., GitHub Copilot, Codex~\cite{chen2021codex} and AlphaCode~\cite{li2022alphacode}), there is also a lot of work exploring the few/zero-shot performance of these models on specific tasks and domains, such as program repair~\cite{prenner2021automatic,kolak2022patch,pearce2022examining}, software security~\cite{asare2022github,pearce2022asleep} and program synthesis~\cite{austin2021program}.

All things considered, in the field of software intelligence, there is no systematic work to evaluate and explore the cross-task generalizability of code models. Therefore, in order to evaluate the cross-task capability of the model on code-related tasks in detail and comprehensively, we build a large-scale benchmark called CrossCodeBench. We start by collecting as many and as diverse code-related tasks as possible, and end up with 216 tasks across 28 categories, 7 types and 18 programming languages. Then, to make our benchmark available for multiple cross-task learning settings (e.g., few-shot learning, learning from task instructions), we manually label each of the 216 tasks with extensive meta-information such as task description, definition, positive/negative examples, etc. Next, we create 10 training/evaluation splits corresponding to different benchmark types.

Given the data splits, we perform experiments by adopting two types of pre-trained models: (1) off-the-shelf model, which are used directly on the evaluation set, and (2) models further fine-tuned on the training set. All models are applied to all splits by using all or suitable cross-task learning methods, such as few-shot learning, learning from instructions, etc. Last but not least, we carry out a number of scaling experiments, through which we find that (1) when the data of each task reaches a certain level (e.g., 10,000 instances), it is difficult for the model to maintain a high improving speed of performance as data instances increasing, and (2) larger models always lead to better performance. We hope that the benchmark, experimental results and analysis we provide will facilitate future research to more powerful cross-task approaches in SE literature. Furthermore, since our benchmark contains massive datasets (and is open to updates), not only our CrossCodeBench, but we hope that such a large-scale meta-dataset (i.e., dataset of datasets~\cite{triantafillou2019meta,wang2022benchmarking}) will facilitate the construction of more benchmarks\footnote{All datasets, tasks and their summaries are available at \url{https://doi.org/10.5281/zenodo.7321934}. Source code is available at \url{https://github.com/NougatCA/CrossCodeBench}.}.

\section{Related Work}
\label{section:related}

\subsection{Few-Shot, One-Shot, and Zero-Shot Learning}
\label{section:related_few_shot}

Albeit defeating human in many fields~\cite{taigman2014deepface,silver2016alphago,he2016resnet,najberg2018alibaba,berner2019dota,brown2020gpt3}, current artificial intelligence (AI) techniques still rely on learning from large-scale task-specific data, and they are unable to rapidly generalize from a few examples. Rather, humans are able to learn new tasks quickly by using what they are born with, or what they have learned in the past.

Few-shot learning (FSL) is therefore proposed in order to learn from a limited number of examples with supervised information. In the cross-task setup, it is an in-context learning approach where a pre-trained language model does not need any fine-tuning and weight updating~\cite{xie2021explanation}\footnote{Since our work only discusses the cross-task scenario, i.e., where no supervised training is performed on the data of the target task, we only introduce the FSL methods that can be applied in this scenario, for more FSL methods please refer to Wang et al.~\cite{wang2020generalizing} and Yin~\cite{yin2020meta}.}. The input can be divided into three parts, namely task description, examples, and prompt~\cite{brown2020gpt3}. The task description is a typically short natural language description of the task, e.g. ``translate English to French''. Examples consist of $k$ canonical supervised examples (usually $10<k<100$, one-shot learning for $k=1$~\cite{brown2020gpt3}), each of which includes the context (i.e., the input/question of the example) and the desired completion (i.e., the output/answer of the example). And the prompt is the context part of the example for which the model needs to make a prediction.

FSL shows promising results compared to the supervised approaches. In the FSL settings, GPT-3~\cite{brown2020gpt3}, a large pre-trained language model, significantly improves the state-of-the-art (SOTA) on various datasets across many task types, such as completion~\cite{paperno2016lambada}, open-domain question answering~\cite{joshi2017triviaqa} and translation~\cite{durrani2014edinburgh}. In addition, Madotto et al.~\cite{madotto2020language} demonstrate that in some task-oriented dialogue system tasks, language model priming FSL is able to achieve similar or better results than fine-tuning-based baseline. Chen et al.~\cite{chen2020few} show that in the FSL settings, a language model can achieve very reasonable performances and outperforms the strongest baseline by an average of over 8.0 BLEU points improvement, across multiple domains. Under the multilingual translation setting, Winata et al.~\cite{winata2021language} find the in-context few-shot cross-lingual prediction results of language models are comparative to the existing SOTA cross-lingual and translation models. In order to investigate whether and how cross-task generalization ability can be acquired, Ye et al.~\cite{ye2021crossfit} propose CrossFit challenge, a task setup that standardizes the training pipeline, data access and evaluation protocol. As a complement, they present the NLP Few-shot Gym, a repository of 160 diverse few-shot NLP tasks. Experimental results show that the cross-task generalization ability can be obtained by using multi-task and meta learning, and can be affected by the selection of seen tasks.

In contrast to FSL, zero-shot learning (ZSL) is proposed to enable generalization to the target task without any examples. Alternatively stated, the language model is given only the task description and the prompt. As only a short task description is required, ZSL is able to provide maximum convenience, potential for robustness, and avoidance of spurious correlations for transferring language models to new tasks~\cite{brown2020gpt3}. However, this presents huge challenges, such as the possibility of ambiguity in the task description, the absence of examples of output formats, etc. Even so, on some tasks, zero-shot GPT-3 can still outperform one-shot GPT-3 (e.g., completion~\cite{paperno2016lambada,zellers2019hellaswag}), or even few-shot GPT-3 (common sense reasoning~\cite{lai2017race}). Besides, on ANLI~\cite{nie2020anli}, GPT-3 under the zero-shot setting scores higher than under the few-shot and one-shot settings for some parameter size settings (i.e., 0.1B, 2.6B and 6.7B).

\subsection{Learning from Task Instructions}
\label{section:related_instructions}

In addition to investigating on the number of examples, researchers do some exploration on task description as well. Recall that in the FSL and ZSL settings, task descriptions are often short (e.g., 12.6 tokens on average~\cite{mishra2022cross}), sometimes causing ambiguity, sometimes missing necessary formatting instructions, etc. Therefore, researchers investigated the use of longer and more detailed task descriptions.

Inspired by current NLP datasets built using crowdsourcing, Efrat and Levy~\cite{efrat2020turking} examine if language models can follow crowdsourcing instructions with no further training. Weller et al.~\cite{weller2020learning} construct a crowdsourced dataset, called Zest, with question-like task descriptions. To study the ability of a model that learns a new task by understanding the human-readable instructions that define it, Mishra et al.~\cite{mishra2022cross} introduce a dataset including 61 distinct tasks, their human-authored instructions and instances, which is named NatInst. Then they adopt existing pre-trained language models to encode task-specific instructions along with input and generate task output. Results show that the model can benefit from instructions and can improve performance by 19\% when evaluated in generalization for unseen tasks. PromptSource~\cite{bach2022promptsource}, FLAN~\cite{sanh2021multitask}, and InstructGPT~\cite{ouyang2022gpt-instruct} also study the cross-task generalization ability by following the provided in-context task instructions. As a subsequent work to NatInst, Wange et al.~\cite{wang2022benchmarking} introduce NatInst$_{v2}$, a benchmark of over 1,600 diverse language tasks and their expert-written instructions, which covers 70+ distinct task types. They also propose T$k$-Instruct, a model trained to follow a variety of in-context instructions, which include plain language task definitions and $k$-shot examples.

The above methods can be collectively referred to as learning from task instructions (LTI). Although the above work differs in the content of task instructions, after reading a lot of recent related work, we believe that a typical task instruction can generally be divided into the following parts: task definition, positive (and negative) examples, and other elements (e.g., ``Things to Avoid''~\cite{mishra2022cross}). Task definition is a detailed definition of the task, unlike task description in FSL, it details how to map the given input to the required output in the current task. As a result, task description is typically longer and more detailed than the task description in FSL\footnote{For example, the task definition of the task description ``Translate English to French'' is ``Given a sentence in English, provide an equivalent paraphrased translation in French that retains the same meaning both through the translation and the paraphrase.''~\cite{wang2022benchmarking}}. In addition, task instruction sometimes contains negative instances as opposed to only positive instances in FSL.

By learning in-context task instructions, models such as InstructGPT and NatInst$_{v2}$ are able to achieve better cross-task generalization performance on unseen tasks than current FSL-based models. For example, on the FLAN benchmark, InstructGPT has about 76.2\% win rate compared to baselines such as FSL-based GPT-3~\cite{brown2020gpt3}, and the 3B-parameter T$k$-Instruct outperforms 175B-parameter InstructGPT by 3.3 ROUGE-L points~\cite{lin2002rouge} when evaluated on 119 unseen tasks.

\subsection{Code-Related Multi-Task Benchmarks}
\label{section:related_benchmark}

There are existing benchmarks that span multiple code-related tasks. CodeXGLUE~\cite{lu2021codexglue}, a benchmark which includes 14 datasets for 10 diversified code-related tasks covering code-to-code, text-to-code, code-to-text and text-to-text scenarios. CodeXGLUE is now widely used to evaluate the performance of pre-trained models of source code on various downstream tasks~\cite{wang2021codet5,guo2022unixcoder}. Elnaggar et al.~\cite{elnaggar2021codetrans} collect a benchmark that contains 6 code-related tasks across 9 programming languages (PLs). Compared to these benchmarks, our work provides a larger scale and more diverse code-related tasks (216 tasks and more than 54M data instances in total). In addition, we also provide systematic and reasonable cross-task task splits, which can comprehensively evaluate cross-task capabilities in various scenarios.

XLCoST~\cite{zhu2022xlcost}, a benchmark for cross-lingual code intelligence proposed by Zhu et al., consists of fine-grained parallel data from 7 PLs and English. This parallel data in a total of 8 languages allow XLCoST to support 10 cross-lingual code-related tasks, for example, program synthesis, code summarization, cross-language code retrieval, etc. Puri et al.~\cite{puri2021codenet} introduce a large-scale dataset CodeNet, aiming to benchmark a variety of critical coding tasks, including code similarity and classification, code translation between a large variety of PLs, and code performance (runtime and memory) improvement techniques. Although these efforts provide large-scale multi-task benchmark, the data for each of their tasks are extracted from the same collected dataset. Therefore, the data distribution between their different tasks is identical, and it is difficult for us to obtain rigorous and valid cross-task evaluation results by these benchmarks.

\begin{table}[t!]
    \centering
    \caption{Comparison of CrossCodeBench and Other Code-Related Benchmarks}
    \label{table:benchmark_compare}
    \resizebox{\linewidth}{!}{%
    \begin{NiceTabular}{lccccc}
        \CodeBefore
            \rowcolors{1}{}{gray!15}
        \Body
            \toprule
                \textbf{Benchmark}    & CrossCodeBench & CodeXGLUE   & CodeTrans   & XLCoST      & CodeNet    \\
            \midrule
                Has description?      & \greencheck    & \redx       & \redx       & \redx       & \redx      \\
                Has definition?       & \greencheck    & \redx       & \redx       & \redx       & \redx      \\
                Has examples?         & \greencheck    & \redx       & \redx       & \redx       & \redx      \\
                Is off-the-shelf?     & \greencheck    & \greencheck & \greencheck & \greencheck & \redx      \\
            \hline
                \# of tasks           & 216            & 38          & 13          & 112         & -          \\
                \# of types           & 7/10           & 6+1/7+1     & 2/3         & 3+1/3+1     & -          \\
                \# of categories      & 28             & 14          & 4           & 5           & -          \\
                \# of datasets        & 66             & 13          & 6           & 1           & 1          \\
                \# of PLs             & 18             & 9           & 9           & 7           & 55         \\
                \# of instances       & 54M            & 6.53M       & 9.32M       & 3.38M       & 13.9M      \\
            \bottomrule
    \end{NiceTabular}
    }%
\end{table}

Table~\ref{table:benchmark_compare} compares our proposed CrossCodeBench with aforementioned benchmarks. We first compare whether each task in these benchmarks has a well-validated task description, definition, positive and negative examples, and whether it is off-the-shelf (i.e., the inputs and outputs are already preprocessed for each task). In addition, we compare their number of tasks, the number of task types (cf. Table~\ref{table:task_type_statistics}, where the first number indicates the number of types in a total of 7 types, the next number indicates the types after considering sub-types, and ``+1'' indicates that there are task types other than those in Table~\ref{table:task_type_statistics}\footnote{CodeXGLUE and XLCoST have two retrieval tasks, namely natural language code search and code-to-code retrieval. In this work, we exclude the retrieval task type because our goal is to evaluate the cross-task capability of the unified model, however, retrieval tasks cannot be converted to a unified text-to-text form.}), the number of datasets included, the number of programming languages (PLs), and the approximate number of total instances.

\section{CrossCodeBench}
\label{section:benchmark}

\subsection{Collecting Tasks}
\label{section:benchmark_collect}

In order to collect as many code-related datasets as possible, we hired 5 Ph.D. student (including one author of this paper) and 6 M.S. students (whose research area is software intelligence) to provide all datasets they have used. Their research span the areas of automated program repair, defect localization, code completion, code summarization, code generation, etc. Then, we recursively expand the scope of collecting datasets by reading surveys and based on paper relationships such as citations, related work, etc. In addition, we conduct an exhaustive search on a number of websites widely used to publish or collect datasets, such as GitHub, HuggingFace Datasets, PapersWithCode Datasets, etc. Finally, we end up collecting a total of 66 code-related datasets.


Since some datasets contain multiple subsets (e.g. CodeSearchNet~\cite{husain2020codesearchnet} contains 6 subsets corresponding to different programming languages) or can support multiple tasks (e.g. the XLCoST dataset~\cite{zhu2022xlcost} can support translation between multiple parallel elements), and they do not correspond to the same task descriptions, we split these datasets into different tasks. As a result, we end up with 216 tasks. Moreover, for the task instances, we directly use the input and output in the form of text sequences provided by the dataset without any further processing. Since we are proposing a unified benchmark for all types of tasks, we formalize the inputs and outputs of all tasks in text-to-text form with reference to T5~\cite{raffel2020t5}. In particular, for the classification task, we convert all labels to corresponding task-relevant text\footnote{For example, for some binary classification tasks, we convert the label ``0'' to text ``No'' and ``1'' to ``Yes''.}. 

\subsection{Meta Information}
\label{section:benchmark_meta}

Some basic information is already done when collecting datasets and tasks, such as the URL, BibTeX, and input/output languages. Next, in order to make our benchmark support multiple cross-task in-context learning methods, namely FSL, ZSL, and LTI, we need to pair each task with some corresponding meta information. In this section, we first present the schema of the meta information we need to complete, then illustrate the open coding procedure~\cite{sgier2012qualitative} that we follow to complete the meta information.

\subsubsection{Schema}
\label{section:benchmark_meta_schema}

With Sections~\ref{section:related_few_shot} and \ref{section:related_instructions}, we decide to match meta information for each task in the following fields:

(a) \textbf{Type}: the type of task to which the task belongs.

(b) \textbf{Description}: a short description of the task, e.g. ``Translate English to French'', ``Summarize'', etc.

(c) \textbf{Definition}: detailed instruction of the task, including a description of the input, and how to map input to output.

(d) \textbf{Positive/Negative Examples}: canonical positive or negative examples.

With these fields, our benchmark can support all current cross-task learning approaches. For FSL, we use task descriptions and positive examples; for ZSL, we only use task descriptions; and for LTI, we use task definitions and positive/negative examples.

\subsubsection{Coding Procedure}
\label{section:benchmark_meta_coding}

Open coding procedure is a widely used standard analytical process that can be utilized to label a dataset~\cite{chaparro2017detecting,zhai2020cpc,chen2021my}. We invite 1 Ph.D. student and 4 M.S. students mentioned in Section~\ref{section:benchmark_collect} to participate in the completion of the meta information, all of whom have more than 4 years of programming experience and have been working on code-related tasks for more than 2 years. As a prerequisite background knowledge, we ask all coders to first read the NLP-related work mentioned in Sections~\ref{section:related_few_shot} and \ref{section:related_instructions}, and to become familiar with the corresponding data as well. Below we show the steps of the coding procedure.

\textbf{(a) \textit{Pilot Study}}. 
With the intent of defining the coding framework, one of the coders conduct a pilot study on randomly selected NLP 30 tasks with meta information, which are equally derived from the work of Brown et al.~\cite{brown2020gpt3}, Bach et al.~\cite{bach2022promptsource}, and Wang et al.~\cite{wang2022benchmarking}. The goal of this study is to identify the initial task type, analyze and identify the common pattern of the task description, definition, and the positive/negative examples. The work ends with the definition of four initial classifications, in addition to the experience and specifications for writing task descriptions (14 patterns), task definitions (5 patterns), and positive/negative examples. Here are details:

\textbf{$\bullet$} Initial task type includes \textit{classification}, \textit{translation}, \textit{generation}, \textit{summarization} and \textit{type prediction}.

\textbf{$\bullet$} Task description is a short single sentence command that often appears before the input instance and connects it to the instructions. It briefly describes the intent of the task, usually with one or a few words. 14 initial patterns are identified, to name a few, \textit{Translate A to B}, \textit{Summarize A}, \textit{Detect (defect/clone/variable misuse)}.

\textbf{$\bullet$} Task definition is a detailed guide to solving the task and can be divided into three parts: (1) the description and interpretation of the input under the task, including the language and form of the input, e.g., \textit{given a Java method with the method name masked by a special symbol `[MASK]'}, \textit{given a natural language description}; (2) how to convert the given input into output, and the language and form of the output, e.g., \textit{translate the given Java method to Python function with the same functionality}, etc.; (3) others, including the search space of the output (for classification tasks, e.g., \textit{if ..., outputs `Yes', otherwise outputs `No'}), and the format of the output (e.g., \textit{for each identifier, outputs the name of the identifier and its type, separated by a colon `:'}). 5 initial patterns are identified which connect these three elements into a complete, fluent, easy-to-understand natural language sentence.

\textbf{$\bullet$} Four positive and four negative examples are required. Examples need to reflect the most typical situation in the task, explaining the main points of generating positive examples for positive examples, and pointing out the errors and giving the correct modifications for negative examples.

Based on this, this coder organizes a 60-minute session for the other four coders for training and discussion.

\textbf{(b) \textit{Completion Procedure}}. 
Each coder is assigned to all 216 tasks. For each task, coders are asked to learn about the task by referring to the websites and paper through the URL and BibTeX, then observe the data for that task, and finally identify the task type and fill in the rest of the meta information as required. 

Specifically, for task types, coders are allowed to identify a task as a new task type despite the fact that an initial task type already exists. In addition, it is encouraged to identify the task type while also giving the sub-types of the task under this task type, if possible. All other coders are informed when a new task type/sub-type is proposed, and coders will have a discussion for at least 10 minutes. If all the coders agree on the new task type/sub-type, then the type/sub-type list would be updated to include the new one.

As for the task description and definition, coders are asked to try to follow the 5/14 initial patterns for writing, and if this is not possible, adding new patterns is also allowed. An online list of writing pattern is shared between all coders, including both the initial schema and the newly added schema. In this way, the latest schema list is visible to all coders and open for them to use and add. It is worth noting that a new pattern can only be added to the list after it has been verified by all coders and the disagreement is solved by discussion.

In addition, for task types/sub-types, pattern lists of task descriptions and definitions, when adding new entries (i.e., new types/sub-types or task description/definition patterns), the similarity between the new entry and the existing entry is inspected. Similar entries are merged into a new, more general entry in due course, without disagreement, and the existing labels are updated accordingly as well. The process is fully iterative and includes continuous refinement of the entry and discussion of ambiguous cases. Each decision made during the entry extraction process represents the opinion of all coders.

Lastly, when it comes to positive/negative examples we are not able to follow the procedure described above since they are all free-form labeled content. We first ask all coders to independently complete writing the task description, task definition, and 2 positive/negative examples. After finishing, we gather all coders to discuss each task, and for the task description and definition, we put together what all coders had written, vote on the most concise and clear version, and revised it until everyone is satisfied. For the positive and negative examples, we first put together a total of 10 examples they written, vote on the most representative 4 until no disagreement remains.

\textbf{(c) \textit{Agreement Measurement}}. 
To evaluate the validity and reliability of our coding procedure, we use Cohen's Kappa value~\cite{cohen1960coefficient} to measure the agreement among all coders. The results are 85.4\%, 76.7\%, 73.2\% and 70.9\% for task type, task description, task definition, selection of positive and negative examples, respectively. This shows that all coders have ``almost perfect agreement'' on the task type and ``substantial agreement'' on the other three meta information~\cite{viera2005understanding}.

For fields that still have disagreements, we resolve them by applying a third person solution. We assign conflicting reports to 4 external coders (the 5 Ph.D. students mentioned in Section~\ref{section:benchmark_collect} except the one internal coder) and let them judge and resolve these disagreements. Our analysis shows that the disagreements mainly focus on the second part of the task definition, where different coders have different understandings and representations of the input-to-output mapping, as well as positive and negative examples.

Finally, in order to ensure the resulting meta information is sufficient for average software engineering researchers. We invite five undergraduate software engineering students with programming experience in non-software intelligence fields to read the meta information of all tasks. They are then asked to review the definition, intent, input and output formats, and other information for each task. Where they are unclear or incorrectly stated, we will feed back to the 4 external coders mentioned in the previous paragraph for revision until the task can be clearly defined and understood.

\subsubsection{Summary}
\label{section:benchmark_meta_summary}

In the end, we get 7 task types as follows.

\textbf{$\bullet$ Classification}: output the corresponding labels based on the input. It is further divided into two sub-types, Binary and Multi-label, which correspond to tasks with only two labels and more than two labels, respectively.

\textbf{$\bullet$ Fill in the blank}: predict the missing token or sequence in a given input.

\textbf{$\bullet$ Translation}: translate code snippets written in one language into another preserving semantics and functionality.

\textbf{$\bullet$ Generation}: generate a sequence based on the input. Three sub-types are (1) Rewrite: modify a part of the given code sequence and output the modified version; (2) Text-to-Code: input the natural language description and output the corresponding code sequence; (3) Code-to-Text: input as code, output the required natural language sequence as required.

\textbf{$\bullet$ Summarization}: given a piece of code, output the functional description of that piece of code\footnote{We separate this task from the Code-to-Text sub-type in Generation because this type of task is in a very important position in software intelligence research. And its research approach is different from other Code-to-Text approaches (e.g., commit message generation)}.

\textbf{$\bullet$ Type Prediction}: predict the type of all identifiers in a given code snippet, which is a kind of sequence tagging tasks.

\textbf{$\bullet$ Question Answering}: given a piece of code and a natural language question, output the answer to that question. 

\begin{table}[t!]
    \centering
    \small
    \caption{Statistics of CrossCodeBench}
    \label{table:benchmark_statistics}
    \begin{NiceTabular}{lr}
        \CodeBefore
            \rowcolors{1}{}{gray!15}
        \Body
            \toprule
                \textbf{Field}                       & \textbf{Average} \\
            \midrule
                \# of instances per task             & 250032.48        \\
                Task description length              & 3.52             \\
                Task definition length               & 26.39            \\
                Positive example input length        & 21.42      \\
                Positive example output length       & 15.77      \\
                Negative example input length        & 21.34      \\
                Negative example output length       & 15.48      \\
            \bottomrule
    \end{NiceTabular}
\end{table}
\begin{table}[t!]
    \centering
    \caption{Task Types and Their Statistics}
    \label{table:task_type_statistics}
    \resizebox{\linewidth}{!}{%
    \begin{NiceTabular}{lrrr}
        \CodeBefore
            \rowcolors{1}{}{gray!15}
        \Body
            \toprule
                \textbf{Type} &
                \textbf{\# of Tasks} &
                \textbf{\# of Categories} &
                \textbf{\# of Instances} \\
            \midrule
                \textbf{Classification}     & \textbf{21}  & \textbf{15} & \textbf{6,565,637}  \\
                \quad - Binary              & 17           & 13          & 6,430,804           \\
                \quad - Multi-label         & 4            & 2           & 134,833             \\
                \textbf{Fill in the Blank}  & \textbf{10}  & \textbf{3}  & \textbf{13,413,614} \\
                \textbf{Translation}        & \textbf{94}  & \textbf{1}  & \textbf{2,366,970}  \\
                \textbf{Generation}         & \textbf{53}  & \textbf{5}  & \textbf{19,511,282} \\
                \quad - Rewrite             & 10           & 2           & 3,197,080           \\
                \quad - Text-to-Code        & 41           & 2           & 15,658,171          \\
                \quad - Code-to-Text        & 2            & 1           & 656,031             \\
                \textbf{Summarization}      & \textbf{34}  & \textbf{2}  & \textbf{11,186,611} \\
                \textbf{Type Prediction}    & \textbf{2}   & \textbf{1}  & \textbf{773,038}    \\
                \textbf{Question Answering} & \textbf{2}   & \textbf{1}  & \textbf{189,863}    \\
            \midrule
                \textbf{Total}              & \textbf{216} & \textbf{28} & \textbf{54,007,015} \\
            \bottomrule
    \end{NiceTabular}
    }%
\end{table}

Table~\ref{table:benchmark_statistics} provides statistical information about CrossCodeBench. In addition, Table~\ref{table:task_type_statistics} shows the statistics of the final task types. In total, the benchmark includes 216 tasks, 28 task categories and over 54M instances.

We define a task \textbf{category} as the set of tasks that have the same intent. 
We define a \textbf{task} as a \textit{\textless task category, dataset\textgreater} pair, and a task \textbf{type} depends on the input and output of the task, not on the intent. A task category is a superset of a task. But a task category is equal to a task if and only if the task category has only one dataset. A task category is a subset of a task type (e.g. Bug Fixing~\cite{tufano2019bfp} and Mutant Generation~\cite{tufano2019mutant} categories are both Rewrite sub-types), they are equal if and only if the task type has only one task category, e.g. the Translation type has only one category, Code Translation.

\subsubsection{Demonstration}
\label{section:benchmark_meta_demonstation}

In our benchmark, each task consists of two json files, one containing the meta information and the other containing the data instances. Both files have the same name in the first part, i.e., ``task\_\{task\_id\}\_\{dataset\_name\}\_\{task\_type\}", followed by ".meta.json" and ".data.json", respectively.

To better illustrate, Figure~\ref{figure:meta_example} shows the contents of a json file with the meta information of a task. We can see that in addition to the fields we mentioned in Section~\ref{section:benchmark_meta_schema}, we also include some information about the task/dataset, such as the language of input/output, BibTeX and URL of the dataset, etc.

\begin{figure}
    \centering
\begin{tcolorbox}[title=task\_006\_swapped\_operands\_classification.meta.json,fontupper=\footnotesize,left=1mm,right=1mm,top=1mm,bottom=1mm]
    ``\textbf{Type}'': [``Classification'', ``Binary''],\\
    ``\textbf{Description}": ``Detect swapped operands'',\\
    ``\textbf{Definition}'': ``You are given a function, your task is to identify whether the operands of non-commutative binary operators are swapped. Construct an answer that is `Swapped operands' if such a swap occurs and `Correct' otherwise.'',\\
    ``\textbf{Input Language}'': ``Programming Language -\textgreater~Python'',\\
    ``\textbf{Output Language}'': ``Natural Language -\textgreater~English'',\\
    ``\textbf{Positive Examples}'': [\{
    
        ~~~~``Input": ``def \_\_contains\_\_(self, x): return x in self.columns'',
        
        ~~~~``Output": ``Correct'',
        
        ~~~~...\\
    ],\\
    ``\textbf{Negative Examples}'': [...],\\
    ``\textbf{BibTeX}'': ...,\\
    ``\textbf{URL}'': ...,\\
    ...
\end{tcolorbox}
\vspace{-0.5em}
    \caption{An example of the task meta information.}
    \label{figure:meta_example}
\end{figure}

\subsection{Splits}
\label{section:benchmark_splits}

After preparing the data, we need to create different training/evaluation splits to evaluate the cross-task learning ability of the model in different application scenarios and difficulties. Table~\ref{table:task_splits} lists the 10 training/evaluation splits we create and use in this paper. We classify them in two dimensions: the level of the cross-task, and the scope of the training task. In addition, we list the name of each split, the number of tasks and instances of the training/evaluation split in Table~\ref{table:task_splits}. Next, we will introduce these splits in order from 3 cross-task levels, namely, cross-category, cross-sub-type and cross-type.

\begin{table}[t!]
    \centering
    \caption{Task Splits and Their Statistics}
    \label{table:task_splits}
    \resizebox{\linewidth}{!}{%
    \begin{NiceTabular}{lllll}
            \toprule
                \Block{2-1}{\textbf{Cross Type}} &
                \Block{2-1}{\textbf{Scope}} & 
                \Block{2-1}{\textbf{Split Name}} & 
                \textbf{\# of Tasks} &
                \textbf{\# of Instances}\\
            &
                &
                &
                \Block{1-2}{\textbf{Train/Eval}} &
                \\
            \midrule
                \Block{4-1}{Cross-Category} &
                \Block{2-1}{Intra-Type} &
                Cat-Intra-CD &
                20/1 &
                4.83M/1.73K \\
            &
                & 
                Cat-Intra-BF &
                44/9 &
                16.4M/3.1M \\
            \cline{2-5}
                & 
                \Block{2-1}{Inter-Type} &
                Cat-Inter-CD &
                215/1 &
                52.3M/1.73M \\
            &
                &
                Cat-Inter-BF &
                207/9 &
                50.9M/3.1M \\
            \hline
                \Block{4-1}{Cross-Sub-Type} &
                \Block{2-1}{Intra-Type} &
                Sub-Intra-ML &
                17/4 &
                6.4M/135K \\
            &
                & 
                Sub-Intra-C2T &
                51/2 &
                18.9M/656K \\
            \cline{2-5}
                & 
                \Block{2-1}{Inter-Type} &
                Sub-Inter-ML &
                212/4 &
                53.9M/135K \\
            &
                &
                Sub-Inter-C2T &
                214/2 &
                53.3M/656K \\
            \hline
                \Block{2-1}{Cross-Type} &
                \Block{2-1}{Inter-Type} &
                Type-Trans &
                122/94 &
                51.6M/2.4M \\
             &
                &
                Type-QA &
                214/2 &
                53.8M/190K \\
            \bottomrule
    \end{NiceTabular}
    }%
\end{table}

\subsubsection{Cross-Category}
\label{section:benchmark_splits_category}
Define new task category is common in SE, but collecting and labeling a large amount of data for the newly defined task category can be very labor-intensive, even impossible. Instead, we can define the task category exactly and give several examples. At this point, we can consider using cross-task learning methods, where learning on a large number of existing tasks acquires generalization capabilities on this new task. Therefore, we first wish to explore whether, and to what extent, current cross-task learning methods will make models achieve cross task category ability. We choose a binary classification task, i.e., clone detection~\cite{svajlenko2014bigclonebench} and a rewrite generation task, bug fixing~\cite{tufano2019bfp} as evaluation task categories, respectively. In addition, since task categories are a subset of task types, we also want to explore whether more data from different task types can help improve the cross-task performance of the model on one task category. 

Therefore, for each of the two evaluation sets, we use two training sets for training: (1) the training set includes only tasks of the same task type as the evaluation task in the other tasks (Intra-Type); (2) the training set includes all other tasks (Inter-Type). Finally, we create 4 splits for the cross-category level, namely Cat-Intra-CD, Cat-Intra-BF, Cat-Inter-CD, and Cat-Inter-BF.

\subsubsection{Cross-Sub-Type}
\label{section:benchmark_splits_sub_type}

Recall that we also have sub-types for Classification and Generation types in Table~\ref{table:task_type_statistics}. We also want to explore to what extent the model can learn cross-task capabilities at the level of sub-type. Therefore, we select two sub-types for each of them, namely Multi-label Classification and Rewrite Generation. For multi-label classification sub-type, we want to investigate whether the model can generalize to multi-label tasks by learning only binary classification sub-type (and other types of tasks). For rewriting tasks, we wish to explore how much rewriting skill the model can acquire without having learned any rewriting task. As with the setup in the cross-category, we create 2 splits for each of the two selected sub-types under the different scopes of the training set. Consequently, we create 4 splits as well, i.e., Sub-Intra-ML, Sub-Intra-C2T, Sub-Inter-ML, and Sub-Inter-C2T.

\subsubsection{Cross-Type}
\label{section:benchmark_splits_type}

The last is the most challenging setup, the cross-type, where we use all tasks in the entire task type as the evaluation set and the other types of tasks as the training set. First, we want to use the Translation type as the evaluation set, and we want to explore whether the model can learn the correspondence between different languages by learning from other types of tasks. Second, we choose a type that contains the fewest tasks, i.e., Question Answering. This split is closest to the real practice situation, i.e., by learning on a well-resourced task type, we expect the model to have a better generalization ability on the new task type without resources. Therefore, we create two splits at the cross-type level, namely Type-Trans and Type-QA.

By now, we create 10 training/evaluation splits of different difficulty levels and different application scenarios. These splits are used in the experiments. Besides, we are able to ensure that there is no data leakage in these splits, in another word, there is no task in the training set with the same resource dataset as any task in the evaluation set. 

\section{Experimental Setup}
\label{section:setup}

In this section we describe the setup for preliminary experiments on CrossCodeBench, including cross-task learning methods, baselines, evaluation metrics and other settings.

\subsection{Cross Task Learning Methods}
\label{section:setup_learning_methods}

Here we introduce the in-context cross-task learning method that we use for our experiments. In-context learning methods also determine the composition of the inputs to the model in our experiments. By referring to Section~\ref{section:related_few_shot}, \ref{section:related_instructions} and the meta information we introduced in Section~\ref{section:benchmark_meta}, we propose the following learning methods.

\textbf{$\bullet$} $k$-shot Learning ($k$-shot) : recall that the input of the few-shot learning consists of three parts: task description, $k$ examples and a prompt. Since each task in CrossCodeBench contains 4 positive examples, we randomly select $k (=4,3,2,1)$ positive examples as the examples of the input context, resulting four methods named \textbf{4-shot}, \textbf{3-shot}, \textbf{2-shot} and \textbf{1-shot}.

\textbf{$\bullet$} Zero-shot learning (\textbf{zero-shot}): same as $k$-shot, but without the second part of the input, i.e. the example.

\textbf{$\bullet$} Learning from instructions with $m/n$ positive/negative examples ($m/n$-instruct): use instructions items introduced in Wang et al.~\cite{wang2022benchmarking}, with $m/n$ positive/negative examples. Specifically, we select the two combinations that worked best shown by Wang et al.~\cite{wang2022benchmarking} (1) \textbf{2/0-instruct}: $m=2$, $n=0$ when training and $m=4$ when evaluating; (2) \textbf{2/2-instruct}: $m=2$ and $n=2$ for both training and evaluating.

Finally, we derive 7 cross-task learning methods that will be used as input to the baseline model to be presented below.

\subsection{Models}
\label{section:setup_models}

\subsubsection{Shortcut Methods}
\label{section:setup_models_shortcut}

The feasibility of a proposed benchmark is important for the subsequent work, which determines whether the benchmark is practical and meaningful to study. Specifically for CrossCodeBench, it is whether the model can actually gain cross-task capability by using a particular cross-task learning method, rather than just behaving like it has gained such capability through some easy shortcuts. Therefore, we propose two such shortcut methods for each split, (1) \textbf{Copy Ex-output}: randomly copying the output of one of the four positive examples and (2) \textbf{Copy Ins-input}: copying the input of the current instance.
We use these two shortcut methods to evaluate the feasibility of CrossCodeBench, i.e., doing cross-task research in the field of software intelligence.

\subsubsection{Off-the-Shelf Models}
\label{section:setup_models_off}

Off-the-shelf models are those that are evaluated directly on the evaluation set, without any further fine-tuning. We first wish to choose a model that is a ``few-shot learner'', such as GPT-3~\cite{brown2020gpt3}. However, our request to use the GPT-3 API has not been approved. Instead, we turn to \textbf{Tk-Instruct}~\cite{wang2022benchmarking}, a T5-based model that is already trained to follow general language task instructions (including a portion of code-related tasks). Tk-Instruct is shown to have better cross-task performance than few-shot learners, such as GPT-3, and other task instruction learners, such as InstructGPT~\cite{ouyang2022gpt-instruct}. Tk-Instruct is only used under the $m/n$-Instruct learning settings. Specifically, we use the 3B parameter version of the model, and for the learning method in the 2/0-instruct and 2/2-instruct settings, we use the ``Tk-Instruct-3b-def-pos' and ``Tk-Instruct-3b-def-pos-neg'' versions of the model, respectively, in order to achieve the best performance~\cite{wang2022benchmarking}. We use the Tk-Instruct model without any fine-tuning and directly evaluate them on the evaluation set, under the different learning methods.

\subsubsection{Fine-tuned Models}
\label{section:setup_models_tuned}

We evaluate pre-trained models that are fine-tuned on the training set with aforementioned in-context learning method as well. First, we choose two recent and widely used pre-trained models of source code, \textbf{PLBART}~\cite{ahmad2021plbart} and \textbf{CodeT5}~\cite{wang2021codet5}. In particular, PLBART is a sequence-to-sequence model based on BART~\cite{lewis2020bart} and pre-trained on an extensive collection of Java and Python functions and associated NL text. CodeT5 is a T5~\cite{raffel2020t5}-based model that is pre-trained on a large corpus containing 8 programming languages and natural language. We ensure that neither PLBART nor CodeT5 had supervised training on any evaluation task in any splits. We also ensure that there is no overlap between the dataset they use in pre-training and the source dataset of the evaluation set in all splits. Besides, we further fine-tune \textbf{Tk-Instruct}~\cite{wang2022benchmarking} by using the $m/n$-instruct learning method. We hope that further fine-tuning will help Tk-Instruct to adapt to code-specific tasks while understanding generic task instruction. Consequently, we obtain three fine-tuned models, two pre-trained models of source code, PLBART and CodeT5, which will be fine-tuned under all cross-task learning methods. There is also an instruction learner, Tk-Instruct, which will be fine-tuned under only two $m/n$-instruct methods. In order to balance efficiency and effectiveness, we use the official ``large'' version of all three models.

\subsubsection{Supervised Models}
\label{section:setup_models_supervised}

We estimate an upper bound performance of each split by supervised fine-tuning a CodeT5-large model on all task instances (except instances used for evaluation) on all evaluation set. This approach follows the classical ``pre-train then fine-tune'' paradigm, where we use all evaluation tasks in the fine-tuning and do not apply any cross-task learning methods. Since this approach allows the model to see the data of the target task and perform supervised fine-tuning, this approach is theoretically the upper limit of the cross-task approach on the corresponding split.

Finally, we end up with 2 shortcut methods, 2 off-the-shelf models, 3 fine-tuned models and 1 supervised model. All model checkpoints are loaded from the official models published on HuggingFace Hub. 

\subsection{Metrics}
\label{section:setup_metrics}

Since outputs of all tasks are in text form (see Section~\ref{section:benchmark_collect}), so text-specific metrics are used in our experiments. Specifically, for classification tasks whose output is short and limited, we adopt \textbf{Exact Match (EM)} to measures the ratio of the instances for which a model produces exactly the same string as the gold labels. For the other tasks, their output is a longer sequence, so we employ \textbf{BLEU (B.)}~\cite{papineni2002bleu} and \textbf{ROUGE-L (R.L)}~\cite{lin2002rouge}. Both are widely adopted string overlap metrics that measure the similarity of between the text sequence generated by the model and the gold sequence. For all metrics, we report scores under percentages.

\subsection{Others}
\label{section:setup_others}

In order to avoid the data imbalance problem, we limit the number of instances in each training tasks to 10,000. These 10,000 instances are fixed across different running if the number of instances for a task exceeds 10,000. Similarly, to avoid the evaluation results being unevenly affected by the amount of task data and to make the evaluation more efficient, we select a fixed number of 500 evaluation instances from each task.

All experiments are conducted on 4$\times$NIVDIA Tesla V100 32Gs with a total fine-tuning epochs of 3. We tune hyperparameters using grid search. We select learning rate from \{1e-5, 3e-5, 5e-5, 1e-4\}, warm-up steps from \{500, 1000, 2000\}, and batch size per device from \{4, 8, 16\}. As a result, for the fine-tuning phase, we use an initial learning rate of 5e-5 for CodeT5 and PLBART, and 1e-5 for Tk-Instruct, with a batch size per device of 16 and 1000 warmup steps for all. We run each experiments three times using different random seeds and report the mean.

\section{Results and Discussion}
\label{section:results}

In this section, we present and discuss the preliminary experimental results on CrossCodeBench.

\subsection{Overall Results}
\label{section:results_overall}

\begin{table*}[t!]
    \centering
    \caption{Overall Benchmarking Results}
    \label{table:results_overall}
    \resizebox{\linewidth}{!}{%
    \begin{NiceTabular}{lllr|rr|r|rr|r|rr|r|rr|rr|rr}
        \CodeBefore
            \cellcolor{gray!15}{6-2,6-3,6-4,6-5,6-6,6-7,6-8,6-9,6-10,6-11,6-12,6-13,6-14,6-15,6-16,6-17,6-18,6-19}
            \rowcolors{7}{}{gray!15}[cols=3-19]
        \Body
            \toprule
                \Block{4-1}{\textbf{Type}} &
                \Block{4-1}{\textbf{Models}} &
                \Block{4-1}{\textbf{Methods}} &
                \Block{1-6}{\textbf{Cross-Category}} &
                &
                &
                &
                &
                &
                \Block{1-6}{\textbf{Cross-Sub-Type}} &
                &
                &
                &
                &
                &
                \Block{1-4}{\textbf{Cross-Type}} &
                &
                &
                \\
            \cline{4-19}
                &
                &
                &
                \Block{1-3}{\textbf{Intra}} &
                &
                & 
                \Block{1-3}{\textbf{Inter}} &
                &
                & 
                \Block{1-3}{\textbf{Intra}} &
                &
                &
                \Block{1-3}{\textbf{Inter}} &
                &
                &
                &
                &
                &
                \\
            \cline{4-19}
                &
                &
                &
                \Block{}{\textbf{CD}} &
                \Block{1-2}{\textbf{BF}} &
                &
                \Block{}{\textbf{CD}} &
                \Block{1-2}{\textbf{BF}} &
                &
                \Block{}{\textbf{ML}} &
                \Block{1-2}{\textbf{C2T}} &
                &
                \Block{}{\textbf{ML}} &
                \Block{1-2}{\textbf{C2T}} &
                &
                \Block{1-2}{\textbf{Trans}} &
                &
                \Block{1-2}{\textbf{QA}} &
                \\
            \cline{4-19}
                &
                &
                &
                \textbf{EM} &
                \textbf{B.} &
                \textbf{R.L} &
                \textbf{EM} &
                \textbf{B.} &
                \textbf{R.L} &
                \textbf{EM} &
                \textbf{B.} &
                \textbf{R.L} &
                \textbf{EM} &
                \textbf{B}. &
                \textbf{R.L} &
                \textbf{B.} &
                \textbf{R.L} &
                \textbf{B}. &
                \textbf{R.L} \\
            \midrule
                \Block{2-1}{Shortcut} &
                \Block{1-2}{Copy Ex-output} &
                &
                \underline{\textbf{50.00}} &
                10.35 &
                20.24 &
                50.00 &
                10.35 &
                20.24 &
                \underline{\textbf{10.75}} &
                2.52 &
                1.43 &
                10.75 &
                2.52 &
                1.43 &
                3.97 &
                15.95 &
                3.29 &
                2.71 \\
            &
                \Block{1-2}{Copy Ins-input} &
                &
                0.00 &
                40.51 &
                54.59 &
                0.00 &
                40.51 &
                54.59 &
                0.00 &
                1.28 &
                3.55 &
                0.00 &
                1.28 &
                3.55 &
                45.44 &
                60.67 &
                0.82 &
                2.00 \\
            \hline
                \Block{2-1}{Off-the-\\Shelf} &
                \Block{2-1}{Tk-\\Instruct} & 
                2/0-instruct &
                2.96 &
                29.91 &
                33.23 &
                2.96 &
                29.91 &
                33.23 &
                0.12 &
                3.58 &
                3.97 &
                0.12 &
                3.58 &
                3.97 &
                17.31 &
                29.87 &
                9.11 &
                8.94 \\
            &
                &
                2/2-instruct &
                2.22 &
                29.65 &
                32.45 &
                2.22 &
                29.65 &
                32.45 &
                0.11 &
                3.66 &
                3.42 &
                0.11 &
                3.66 &
                3.42 &
                16.46 &
                29.27 &
                8.10 &
                8.38 \\
            \hline
                \Block{16-1}{Fine-\\Tuned} & 
                \Block{7-1}{PLBART} &
                4-shot &
                15.86 &
                44.76 &
                57.52 &
                25.54 &
                55.86 &
                67.89 &
                1.75 &
                7.92 &
                6.61 &
                8.54 &
                11.95 &
                12.84 &
                \underline{50.14} &
                63.84 &
                12.37 &
                14.61 \\
            &
                & 
                3-shot &
                12.94 &
                43.60 &
                56.40 &
                23.49 &
                54.38 &
                65.21 &
                1.23 &
                7.58 &
                6.65 &
                5.82 &
                11.61 &
                11.59 &
                47.90 &
                62.35 &
                11.86 &
                13.59 \\
            &
                & 
                2-shot &
                10.59 &
                40.84 &
                54.37 &
                22.01 &
                49.73 &
                64.79 &
                0.89 &
                5.44 &
                5.54 &
                3.74 &
                10.73 &
                11.12 &
                47.60 &
                60.76 &
                9.61 &
                11.70 \\
            &
                &
                1-shot &
                9.27 &
                36.45 &
                46.90 &
                21.95 &
                45.95 &
                54.78 &
                0.56 &
                5.86 &
                4.03 &
                3.12 &
                10.51 &
                10.78 &
                46.66 &
                59.83 &
                8.84 &
                9.86 \\
            &
                &
                zero-shot &
                7.65 &
                45.36 &
                44.08 &
                16.40 &
                52.04 &
                50.34 &
                0.03 &
                3.66 &
                3.61 &
                2.81 &
                7.63 &
                6.87 &
                41.48 &
                54.31 &
                6.99 &
                8.33 \\
            &
                &
                2/0-instruct &
                19.11 &
                \underline{47.74} &
                \underline{59.23} &
                53.25 &
                \underline{59.19} &
                70.55 &
                4.82 &
                8.18 &
                7.22 &
                14.66 &
                14.31 &
                \underline{15.43} &
                48.88 &
                \underline{65.70} &
                15.72 &
                18.73 \\
            &
                &
                2/2-instruct &
                19.86 &
                42.84 &
                57.74 &
                \underline{53.87} &
                \underline{58.80} &
                68.89 &
                4.76 &
                \underline{8.96} &
                7.82 &
                14.82 &
                14.16 &
                14.24 &
                47.43 &
                63.95 &
                15.37 &
                17.76 \\
            \cline{2-19}
                &
                \Block{7-1}{CodeT5} &
                4-shot &
                15.80 &
                \underline{46.89} &
                \underline{58.15} &
                30.03 &
                56.92 &
                72.01 &
                1.80 &
                7.38 &
                6.80 &
                8.30 &
                12.15 &
                12.04 &
                49.22 &
                \underline{66.26} &
                14.78 &
                15.12 \\
            &
                &
                3-shot &
                13.05 &
                44.67 &
                57.33 &
                29.55 &
                54.30 &
                69.41 &
                1.04 &
                7.84 &
                6.59 &
                6.83 &
                11.49 &
                10.06 &
                47.83 &
                64.35 &
                14.27 &
                14.99 \\
            &
                &
                2-shot &
                12.49 &
                40.94 &
                52.68 &
                29.53 &
                51.26 &
                62.70 &
                0.81 &
                6.88 &
                5.85 &
                3.47 &
                10.50 &
                10.46 &
                48.65 &
                \underline{65.25} &
                13.69 &
                14.73 \\
            &
                &
                1-shot &
                10.68 &
                35.32 &
                45.38 &
                28.75 &
                48.86 &
                58.14 &
                0.34 &
                6.37 &
                5.57 &
                2.78 &
                9.87 &
                \underline{8.47} &
                45.72 &
                62.92 &
                11.41 &
                12.25 \\
            &
                &
                zero-shot &
                7.12 &
                30.69 &
                43.07 &
                18.8 &
                45.85 &
                56.51 &
                0.05 &
                4.15 &
                3.72 &
                1.95 &
                6.44 &
                6.99 &
                43.24 &
                58.38 &
                9.12 &
                10.95 \\
            &
                &
                2/0-instruct &
                20.89 &
                \textbf{48.33} &
                \textbf{61.02} &
                \underline{56.33} &
                \textbf{61.89} &
                \textbf{76.89} &
                5.57 &
                8.35 &
                7.10 &
                \textbf{15.95} &
                \underline{15.40} &
                \underline{15.63} &
                \textbf{52.05} &
                \textbf{67.87} &
                \textbf{17.95} &
                \underline{20.52} \\
            &
                &
                2/2-instruct &
                20.05 &
                45.57 &
                \underline{59.31} &
                \textbf{56.42} &
                58.49 &
                \underline{74.26} &
                5.06 &
                8.05 &
                6.93 &
                15.10 &
                14.60 &
                \textbf{15.90} &
                \underline{51.51} &
                \underline{66.61} &
                \underline{17.31} &
                \underline{20.12} \\
            \cline{2-19}
                &
                \Block{2-1}{Tk-\\Instruct} &
                2/0-instruct &
                18.43 &
                43.86 &
                57.23 &
                51.88 &
                57.45 &
                70.98 &
                4.92 &
                \textbf{9.18} &
                \textbf{8.47} &
                \underline{15.64} &
                \textbf{15.60} &
                14.35 &
                45.49 &
                62.14 &
                17.04 &
                \textbf{20.67} \\
            &
                &
                2/2-instruct &
                18.52 &
                42.01 &
                56.16 &
                51.20 &
                57.67 &
                70.39 &
                4.24 &
                \underline{9.10} &
                7.88 &
                14.43 &
                \underline{15.01} &
                14.77 &
                45.63 &
                62.41 &
                16.15 &
                19.33 \\
            \hline
                \Block{}{Supervised} &
                CodeT5 &
                None &
                78.49 &
                73.41 &
                82.40 &
                78.49 &
                73.41 &
                82.40 &
                63.23 &
                28.11 &
                27.86 &
                63.23 &
                28.11 &
                27.86 &
                72.26 &
                85.83 &
                37.46 &
                39.49 \\
            \bottomrule
    \end{NiceTabular}
    }%
\end{table*}

Table~\ref{table:results_overall} shows the overall benchmarking results. Except scores of the supervised model, the best scores are bolded and those within 5\% below the best scores are underlined, and if there are no scores within 5\% below, the bolded best scores are further underlined. Based on these results, we have the following observations and discussions.

\subsubsection{Room for improvement exists}

First of all, we can see that supervised fine-tuned CodeT5 following the classical ``pre-train then fine-tune'' paradigm has very significant advantages over any other methods. Since supervised CodeT5 performs supervised training on the target task, it can be seen as an upper bound for the performance of large language models of source code on the corresponding evaluation set. This suggests that there is still a lot of room for theoretical improvement in cross-task learning based on pre-trained models of source code.

\subsubsection{Shortcut wins on low-resource classification}

We find that the best fine-tuned model outperforms shortcut methods on all metrics for all splits except Cat-Intra-CD and Sub-Intra-ML. This illustrates that the cross-task learning approach allows the models to gain real cross-task capability, not just some simple shortcuts. And on the two splits where the shortcut method wins, we discover two points, one is that scopes of the training set of both splits are both ``Intra'', and the second is that only the evaluation sets of these two splits contain the classification tasks.

For the first point, recall that in Section~\ref{section:benchmark_splits}, we create splits of ``Intra'' and``Inter'' by changing the scope of tasks in the training set, the former restricts the tasks in the training set to the same type as the verification tasks, while the latter has all other tasks in the training set. On the two ``Inter'' corresponding to these two ``Intra'', the fine-tuned method outperforms the shortcut method instead. Therefore, we can conclude that having more data for other types of tasks could improves the cross-task performance on tasks of a certain type.

As for the second point, we believe that the main reason is that the output of these tasks is short (usually one word) and fixed (corresponding to all labels), and our experiments do not restrict these conditions, relying only on giving some ``soft'' hints in the task definition and in the output of the examples. For example, in the classification task shown in Figure~\ref{figure:meta_example}, we illustrate the restriction/range of output only in two places, (1) ``Definition'': the output are ``Swapped Operands'' in the case of operands swapping only in definition, and ``Correct'' otherwise, (2) ``Positive/Negative Examples''. These restrictions are simply entered into the model as plain text, and thus cannot pose any substantial restrictions on what the model generates. In this case, we can only expect the model to be trained to acquire the knowledge of how to find the output range on each classification task in the input. But the results show that this is far from the case, and the model often generates text outside the output range, leading to a low performance. This conclusion is further corroborated by browsing the output of the fine-tuned model on the classification task.

The performance of the shortcut method on the classification task suggests that further restrictions on the model's output on the classification task are needed if one wants a unified text-to-text model to have excellent cross-task performance on the classification task.

\subsubsection{Further fine-tuning matters}

It is clear that the fine-tuned model has a significant performance advantage over the off-the-shelf model. This is because fine-tuning on our training set can help the model understand the input representation of cross-task learning methods (for PLBART and CodeT5), or become more familiar with tasks in the software intelligence domain (for Tk-Instruct). Even though Tk-Instruct has included several code-related tasks, such as ``code-to-text'', in its original training set~\cite{wang2022benchmarking}, we believe that these tasks are diluted in a huge number of NLP tasks, making off-the-shelf Tk-Instruct unable to obtain good power in code-related tasks.

\subsubsection{Domain-specific models are better}

We also notice that CodeT5, a pre-trained model of source code, is able to achieve better results than Tk-Instruct, a model trained to follow task instructions, in most tasks. This suggests that domain-specific pre-trained models are generally a better choice than cross-domain models. Moreover, we also notice that Tk-Instruct wins in the task of generating natural language, such as Code-to-Text and Question Answering. This is not difficult to understand, because training on huge amount of natural language corpus helps it generate more fluent and reasonable natural language text.

\subsubsection{Detailed in-context information generally helps}

Among all cross-task learning methods, LTI produces the best results for all tasks and metrics. Specifically, in most cases, using only two positive examples is more effective than using two positive and two negative examples. By comparing the input representations of FSL, ZSL and LTI, we can see that compared to FSL and ZSL, LTI extends the short task description to a detailed task definition, which allows the model to understand the task more comprehensively, including the input/output format, the search space of outputs, etc., and thus learn the solution to the task. In addition, LSI further introduces negative examples, but it appears that the extra negative examples do not contribute positively to the performance of the model in most cases. This also confirms the results obtained by Wang et al.~\cite{wang2022benchmarking}.

\subsection{Data Scaling}
\label{section:results_data}

\begin{figure}[t!]
    \centering
    \includegraphics[width=\linewidth]{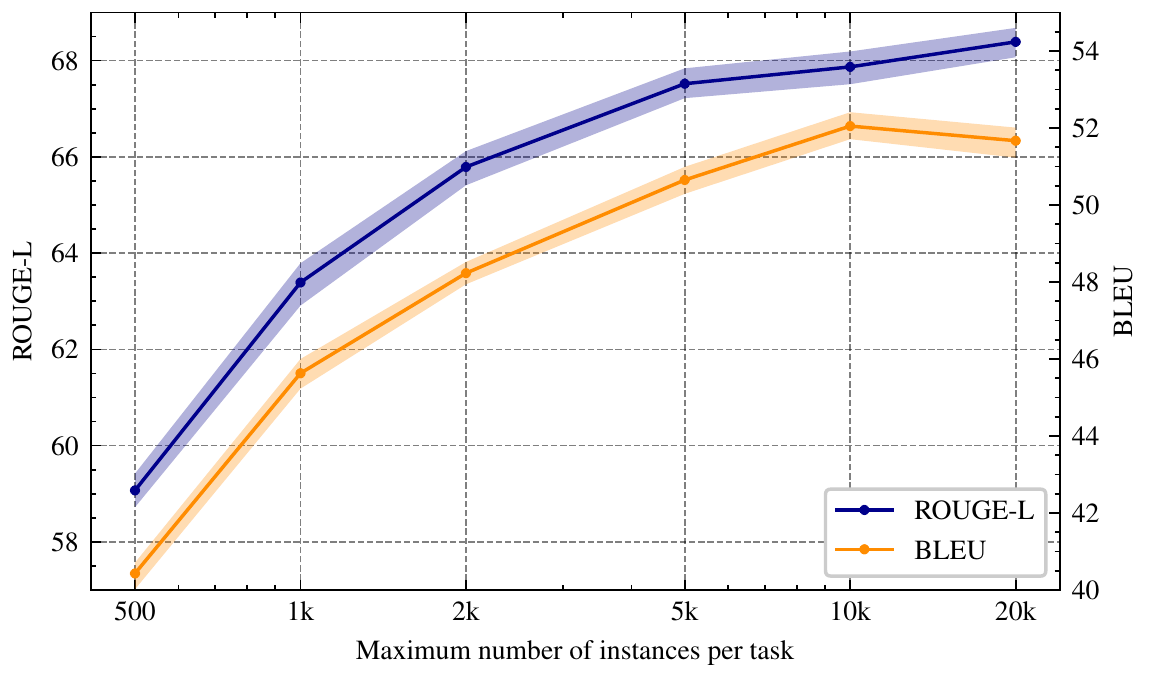}
    \caption{Scaling trends of models performance as the maximum number of instances per task changes from 500 to 20,000. The error is indicated by light shading. The x-axis is on a log-scale in order to be more intuitive.}
    \label{figure:data_scaling}
\end{figure}

In addition to more data on other task types, we also explore whether the size of the training instances could affect the cross-task performance of the model given the same tasks. Therefore, on Type-Trans, we vary the number of instances per task that are used for fine-tuning CodeT5, and the evaluation results are shown in Figure~\ref{figure:data_scaling}.

It can be seen that when the number of tasks in the training set is constant, increasing the maximum number of instances per task within a certain range can significantly improve the cross-task performance of the model. However, when the maximum number of instances per task increases from 10,000 to 20,000, the BLEU shows a significant decrease, so for the BLEU metric, the model achieves its best performance at 10,000. On the other hand, for ROUGE-L, although the average value has been increasing all the time, if the range of data errors is taken into account, the model may start to show a decrease in performance when increasing from 5,000 to 10,000. Thus, in aggregate, the model may achieve its best performance at a maximum number of instances of 10,000 per task, which, if continued to increase, leads to time and space costs that do not match the gains in model performance.

\subsection{Model Scaling}
\label{section:results_model}

\begin{figure}[t!]
    \centering
    \includegraphics[width=\linewidth]{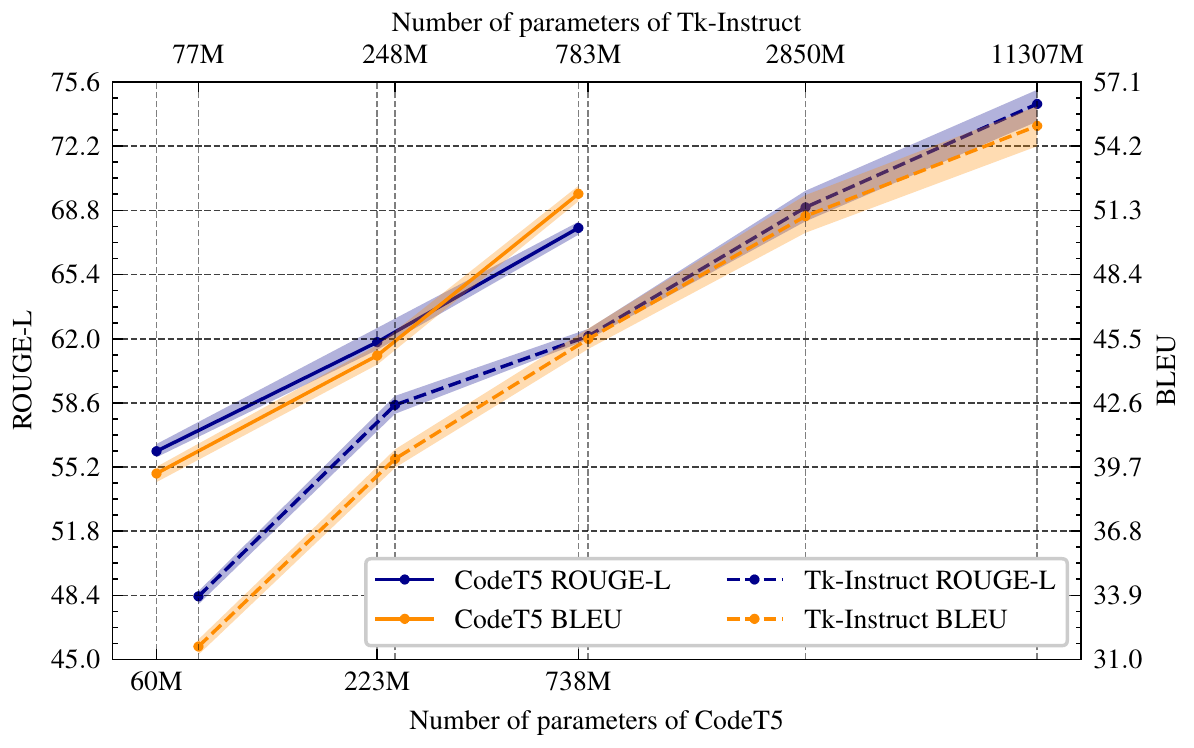}
    \caption{Scaling trends of models performance as the number of model parameter changes. The error is indicated by light shading. The x-axis is on a log-scale in order to be more intuitive.}
    \label{figure:model_scaling}
\end{figure}

We also study the effect of model scaling by initializing CodeT5 and Tk-Instruct from different sizes of checkpoints and the results are in Figure~\ref{figure:model_scaling}.

We find that increasing the model size consistently delivers an improvement in the cross-task performance of the model over the range of model sizes we experimented with, and is roughly log-linear with parameter size. Combining the results in Figure~\ref{figure:data_scaling}, we can see that models of a certain size have a fixed demand on the amount of data. For example, CodeT5-large reaches a performance bottleneck at a maximum of 10k instances per task. Conversely, a larger model always leads to better performance when the data scale is of a certain size (note that the data volume should be large enough to avoid overfitting).

\subsection{Scope Scaling}
\label{section:results_scope}

Finally, we investigate a problem with practical applications, i.e., the case when we propose a new task class and obtaining the corresponding data is very costly or impossible. Considering the use of cross-task learning is a feasible approach at this point. Thus, there is a question, how the scope of the training set is selected is important for the cross-task capability of the model. Therefore, we select a target task category, Bug Fixing~\cite{tufano2019bfp}, and then change various ranges of training sets to investigate under which scope the model would achieve the best cross-task performance for the Bug Fixing task category.

We also present two splits in Section~\ref{section:benchmark_splits_category}, with experiments on two application scenarios where the target task is Bug Fixing. The training sets in both splits in Section~\ref{section:benchmark_splits_category} include the sub-type (Rewrite) or type (Generation), to which Bug Fixing belongs. However, in practical, the proposed task category may be a new sub-type/type, and other task data of the same sub-type and type is not available. What we further study here is whether in this case, the more data is still better when we can only select tasks of other sub-types or types.

After selecting Bug Fixing as the target task category, we change the scope of the training set data from three dimensions: (1) In-Sub-Type: whether the training set contains task data of the same task sub-type as the target task, i.e., Rewrite; (2) In-Type: whether the training set has data of the same task type as the target task, i.e. Generation; (3) Out-Type: whether there are data in the training set that do not belong to the same task type as the target task, i.e., the other 6 types except the Generation type. We investigate the data in the training set under various combinations in these three dimensions and present the results in Table~\ref{table:scope_scaling}.
\begin{table}[t!]
    \centering
    \caption{Cross-Task Performance of the Model on Different Scope of Fine-Tuning Data}
    \label{table:scope_scaling}
    \resizebox{\linewidth}{!}{%
    \begin{NiceTabular}{ccccrrr}
        \CodeBefore
            \rowcolors{3}{}{gray!15}
        \Body
            \toprule
                \Block{2-1}{\textbf{\#}} &
                \Block{1-3}{\textbf{Fine-tuning data comes from}} &
                &
                &
                \Block{2-1}{\textbf{Training}\\ \textbf{size}} &
                \Block{2-1}{\textbf{BLEU}} &
                \Block{2-1}{\textbf{ROUGE-L}} \\
            \cline{2-4}
                &
                \textbf{In-Sub-Type} &
                \textbf{In-Type} &
                \textbf{Out-Type} &
                &
                \\
            \midrule
                \textbf{1} & \greencheck & \redx       & \redx       & 116K  & 43.41 & 55.82 \\
                \textbf{2} & \greencheck & \greencheck & \redx       & 16.4M & 48.33 & 61.02 \\
                \textbf{3} & \greencheck & \greencheck & \greencheck & 50.9M & 61.89 & 76.89 \\
            \hline
                \textbf{4} & \redx       & \greencheck & \redx       & 16.3M & 28.22 & 43.47 \\
                \textbf{5} & \redx       & \greencheck & \greencheck & 50.8M & 57.13 & 74.82 \\
            \hline
                \textbf{6} & \redx       & \redx       & \greencheck & 34.5M & 53.74 & 71.63 \\
            \bottomrule
    \end{NiceTabular}
    }%
\end{table}

Comparing experiments 2 and 4, or 3 and 5 in Table~\ref{table:scope_scaling}, we learn that the data within the sub-type plays an important role. But such a conclusion holds only when other task data are available. If we compare experiment 1 with 5 and 6, it can be found that the performance achieved by using only in-sub-type data is not as good as that achieved by using data outside the sub-type. 

Besides, in comparison with experiments 2 and 6, we can find that the performance using in-type data is worse than that using out-type data only. We believe the reason is that the amount of out-type data is larger than that of in-type data, which allows the model to learn more cross-task capabilities with more in-context information.

Therefore, we can conclude that the amount of data still plays an important role. Its contribution to the cross-task performance of the model is more important than the data for the same type of task.

\subsection{Implications}
\label{section:results_implications}

After our experiments and discussions above, we have some findings and implications that can facilitate subsequent researchers and practitioners.

The first thing we can see is that all models performs very poorly on the classification task in the cross-task setting. After adding the instruction containing the set of classification labels to the input, the performance of the model on the classification task, though improved, is still poor because it still often outputs words outside the output range. So, how to constrain the model to output valid classification labels requires further investigation.

In addition, our results show that the different combinations of SE tasks used for training a model would yield different performances on a target task. So, when given a new SE task that lacks data, how to quickly determine which existing tasks should be used to achieve the best results is an urgent issue to explore.

Finally, we learn from Sections~\ref{section:results_data}--\ref{section:results_scope} that more data, larger model and more tasks can all benefit the cross-task generalization ability. Therefore, if there are limited resources, is it better to devote it to a larger model or to more data, and if to data, is it better to increase the data per task or to increase the diversity of tasks. That is, how to make better use of the limited resources is also an issue worth investigating.

\section{Threats to Validity}
\label{section:threats}

\paragraph*{Internal Validity}

Threats to internal validity relate to the bias introduced when manually annotating some fields in meta information of the task. To mitigate this, we start with the open coding procedure to reduce the impact of individual bias. Then for remaining conflicts, we address these by employing a third person solution (see Section~\ref{section:benchmark_meta_coding} for details). We believe that the approach we adopt minimizes the human impact on the meta information, but it still may not be completely avoidable.
Another factor for threatening internal validity is hyperparameters. Due to time constraints, we only perform a coarse-grained hyperparameter search. Therefore, other hyperparameter settings may lead to better results. But this does not affect the observations and conclusions presented in this paper, since all experiments are carried out with the same hyperparameter searching strategy. 

\paragraph*{External Validity}

Threats to external validity concern about the data imbalanced among different types of tasks. Even though we limit the maximum number of instances per task, the number of tasks contained in different task categories, types varies greatly (see Table~\ref{table:task_type_statistics}). Therefore, the model may have different generalization performance on different types of unseen tasks. In addition, since we need to apply tasks to the unified sequence-to-sequence model, all tasks in our benchmark are formalized in a text-to-text form. Those code-related tasks that cannot be converted to this form are excluded. An example is the retrieval tasks, such as natural language code search and code-to-code retrieval~\cite{guo2022unixcoder}. The retrieval task requires computing the similarity between the representation vectors generated by the model for different inputs and cannot be applied to the text-to-text framework.

\paragraph*{Construct Validity}

The major threat is the data or task overlap between the evaluation set and the data/task the model has seen during self-supervised pre-training and fine-tuning on the training set. First, since the minimum cross-task granularity of our proposed split is task categories, there is no overlap of task categories between the training and evaluation sets. Then, to mitigate such data overlap threats, we use the tool provided by Allamanis~\cite{allamanis2019adverse} to ensure that the data in the evaluation set of the proposed 10 splits do not have duplication with the data in the training set during our experiments. Therefore, we can guarantee that no overlap problem will be introduced in the fine-tuning phase, but since we use pre-trained models as the backbone of our baselines, such threats may be introduced in the pre-training phase.

\section{Conclusion}
\label{section:conclusion}

To study the problem of cross-task generalization in software intelligence, we collect a large-scale meta-dataset containing various types of tasks and label each task with rich meta information to support various cross-task learning methods. We then create a benchmark, CrossCodeBench, for studying the generalization ability of source code-related deep learning models. Through preliminary experiments, we demonstrate and analyze the feasibility and possible research directions for cross-task studies of software intelligence.

\section*{Acknowledgment}

This work was supported by National Natural Science Foundation of China (61802167), Cooperation Fund of Huawei-NJU Creative Laboratory for the Next Programming, and NSF award 2034508. We also thank the reviewers for their helpful comments. Chuanyi Li is the corresponding author.

\bibliographystyle{IEEEtran}
\bibliography{references}

\end{document}